\documentclass[showpacs,preprintnumbers,amsmath,amssymb,aps]{revtex4}

\usepackage{graphicx}
\usepackage{dcolumn}
\usepackage{bm}

\newcommand{\binomial}[2]{\left(\begin{array}{c} #1\\ #2\end{array}\right)}

\begin{document}


\title{Towards a physics of evolution: 
Existence of gales of creative deconstruction in evolving technological networks}

\author{Rudolf Hanel$^{1,2}$}
\author{Stuart A. Kauffman$^{3}$}
\author{Stefan Thurner $^{1,}$}
\email{thurner@univie.ac.at}
\affiliation{
   $^{1}$ Complex Systems Research Group; HNO; 
   Medical University of Vienna; W\"ahringer G\"urtel 18-20; A-1090; Austria \\
   $^{2}$ Institute of Physics; University of Antwerp; Groenenborgerlaan 171; 2020 Antwerp; Belgium\\
   $^{2}$ Institute for Biocomplexity and Informatics; University of Calgary; 
         2500 University Dr. NW; Calgary; AB T2N 1N4; Canada\\
         }
 
\begin{abstract}
Systems evolving according to the standard concept of biological or technological evolution
are often described by catalytic evolution equations. We study the structure of these equations 
and find a deep relationship to classical thermodynamics. In particular we can demonstrate the 
existence of several distinct phases of evolutionary dynamics: a phase of fast growing diversity,
one of stationary, finite diversity, and one of rapidly decaying diversity. While the first two 
phases have been subject to previous work, here we focus on the destructive  aspects 
-- in particular the phase diagram -- of evolutionary dynamics. 
We further propose a dynamical model of diversity which captures spontaneous creation 
and destruction processes fully respecting the phase diagrams of evolutionary systems. 
The emergent timeseries show a Zipf law in the diversity dynamics, which is e.g. observable in 
actual economical data, e.g. in firm bankruptcy data. We believe the present model is a way to 
cast the famous qualitative picture of  Schumpeterian economic evolution, into a quantifiable and 
testable framework.  
\end{abstract}



\maketitle

\section{Introduction}
Simplistically technological evolution is a process of (re)combination and substitution 
of existing elements to invent and produce new goods, products or things. 
New things can come into existence through combining existing ones 
in whole or part. The new things then undergo a 'valuation' (selection) process based on 
their  'utility' associated to them in the context of  their surroundings. 
The surroundings are defined by  all other yet existing things, 
and all things which may come into existence in the foreseeable future.  Another way how new things can 
come to being is pure chance, such as random inventions which do not rely on pre-existing things.
Biological evolution is a special case of technological evolution (i.e. innovation), where recombination and 
substitution happens through sexual reproduction and mutations.

The dynamics of systems capable of evolution have been formalized some time ago.
In this context the concept of the {\em adjacent possible} has been brought forward
\cite{origin}. The adjacent possible is the set of objects that can get produced within a 
given time span into the future. 
What {\em can} get produced in the next timestep depends crucially  on the 
details of the set of elements that exist now.  
To capture the dynamics of an evolving system which is governed by a combination/substitution 
mechanism,  imagine that the 
diversity of the system is given by a $d$ dimensional state vector $x$. Each element 
$x_i$ characterizes the 
abundance of all possible elements $i$. 
This means that the total number of all elements that can potentially ever exist in the system 
are bounded from above by $d$ \footnote{It was shown in \cite{hanel05} that the limit $d\to \infty$ 
exists and is well defined.}. 
Its dynamics is governed by the famous equation
\begin{equation}
\frac{\partial  }{\partial t} x_i = \alpha_{ijk} x_j x_k - x_i \sum_l \alpha_{ljk} x_j x_k  \quad, 
\label{model}
\end{equation}
where the second term ensures normalization of $x$. $x$ thus captures the {\em relative} abundances 
of existing elements.  
The tensor elements $\alpha_{ijk}$ serve as a 'rule table', telling which combination of two elements 
$j$ and $k$ can produce a third (new) element $i$. The element $\alpha_{ijk}$ is the rate at which 
element $i$ can get produced, given the elements  $j$ and $k$ are abundant at their respective 
concentrations $x_j$ and $x_k$. 
Equation (\ref{model})  has a long tradition; some of its special cases are 
the Lotka Volterra replicators see e.g. in \cite{lotka}, the hypercycle \cite{hyper}, or the Turing gas \cite{turing}. 
Equation (\ref{model})  has been analyzed numerically \cite{farmer1,numerical}, 
however system sizes are extremely  limited. 
In contrast to the amount of available qualitative  and historical knowledge on evolution \cite{gould},   
surprisingly little effort has been undertaken to solve Eq. (\ref{model}) explicitly. 
To understand the dynamics of  Eq. (\ref{model}) more deeply and analytically it was 
suggested in \cite{hanel05} to assume three things: (i) the focus is shifted from the actual concentration of elements 
$x_i$,  to the system's diversity. Diversity is defined as the number of existing elements. 
An element exists, if $x_i>0$, and does 
not exist if $x_i=0$. (ii) For simplicity, the rule table $\alpha$ is assumed to have binary entries, $0$ and $1$ 
only, (iii) the location of the non-zero entries is perfectly random. To characterize the number of  these entries 
the number $r$ is introduced, which is the rule table density or the density of {\em productive} pairs.
The total number of productive pairs in the system (i.e. the number of non-zero
entries in  $\alpha$) is consequently given by $r\,d$.

With these assumptions, the idea in \cite{hanel05} was to explicitly formalize the concept of the adjacent possible, so  
that   Eq. (\ref{model}) could be rewritten into a dynamical map whose asymptotic limit could be 
found analytically. The only variable of the corresponding map is $r$. The initial condition, i.e., the 
initial size of 
present elements is assigned $a_0$. The solution of the system is the asymptotical value ($t\to \infty$) of diversity, 
 $a_{\infty }$.
The amazing result of this solution, (as a function of $r$ and the initial condition $a_0$) is that evolutionary systems 
of the type of Eq. (\ref{model}) have a phase transition in the $r$-$a_0$ plane. 
In one of the two phases -- 
after a few iterations -- no more elements can be built up from existing ones and the 
total diversity converges to a finite number (sub-critical phase). The other phase is 
characterized that the advent of new elements creates so many more possibilities to create 
yet other elements that the system ends up producing all or almost all possible $d$ elements 
This we call the super-critical or 'fully populated' phase.
Even though the existence of a phase transition 
was hypothesized some time ago in \cite{origin}, it is entirely surprising that the phase transition is 
mathematically of exactly the same type as a Van der Waals gas \footnote{It is maybe noteworthy that the 
Fisher structure (linear form of Eq. (\ref{model})) does not have such a transition, for this a non-linear 
model is needed. }.
Note  that this model is a mathematically tractable variant of the so called  bit-string model 
of biological evolution, introduced in \cite{origin}. 

The dynamics discussed so far assumes that a system is starting with relatively low 
diversity $a_0$, which increases over time, up to a final  asymptotic level, $a_{\infty }$.  
However, also the opposite dynamics is possible. Imagine one existing element, say $i$, is 
removed from the system, a species is dying out, or a technical tool gets out of fashion/production. 
This removal can imply that other elements, which needed $i$ as a production  input will also 
cease to exist, unless some other way exists to produce them (not involving $i$). 
Note, that all the necessary information is incorporated  in $\alpha$. 

The first part of this paper studies the dynamics of evolutionary systems which exist in the 
highly populated phase, and where $\delta_0$ elements get kicked out at the initial timestep.  
These defected elements may trigger others to default as well. 
We demonstrate  the existence of a new phase transition in the $\delta_0$-$r$ plane, 
meaning that for a fixed rule density $r$ there exists a critical value of initial defects, above which 
the majority of elements will die out in a cascade of secondary defects. 

The understanding  of these phase diagrams  teaches something  about the 
class of dynamical systems to which the mechanism of evolution belongs to. 
However,  this is only part of the story: it does not 
yet constitute the (microscopic) dynamics of the system \footnote{Note an analogy here 
between the similarity of thermodynamics and statistical physics. The knowledge of 
a phase transition of water does not imply an atomistic view of matter.}. 

In reality, the final diversity  $a_{\infty }$ will not be a constant, but will be subject to fluctuations.
The relevant parameter will become the diversity (number of nonzero elements in $x$) over time, 
$a_t$. 
In particular, there are two types of fluctuations: elements will get created spontaneously with a 
given rate,  and existing elements will go extinct with another rate. The second part of this work 
proposes  a dynamical model of an evolutionary system incorporating these spontaneous processes, 
compatible with their inherent phase diagrams. 
The model is characterized by the rule density $r$, one creation and one destruction process, 
the latter ones  modeled by simple Poisson processes. 
We study the resulting dynamics and find several characteristics typical to 
critical systems and destructive economical dynamics described qualitatively by J. A. Schumpeter a 
long time ago \cite{schumpeter11}.

\section{The creative phase transition}

The dynamics of diversity (number of existing elements over time) has 
been analytically solved in  \cite{hanel05}. To be self-consistent in this section we review the argument: 
It is first  assumed that 
the system has a growing mode only (tensor elements $\alpha_{ijk}$ are zero or one but never 
negative). For this situation  
Eq. (\ref{model}) was projected onto a dynamical map, whose asymptotic 
solutions can be found.  

\begin{figure}
\begin{center}
\includegraphics[width=8cm]{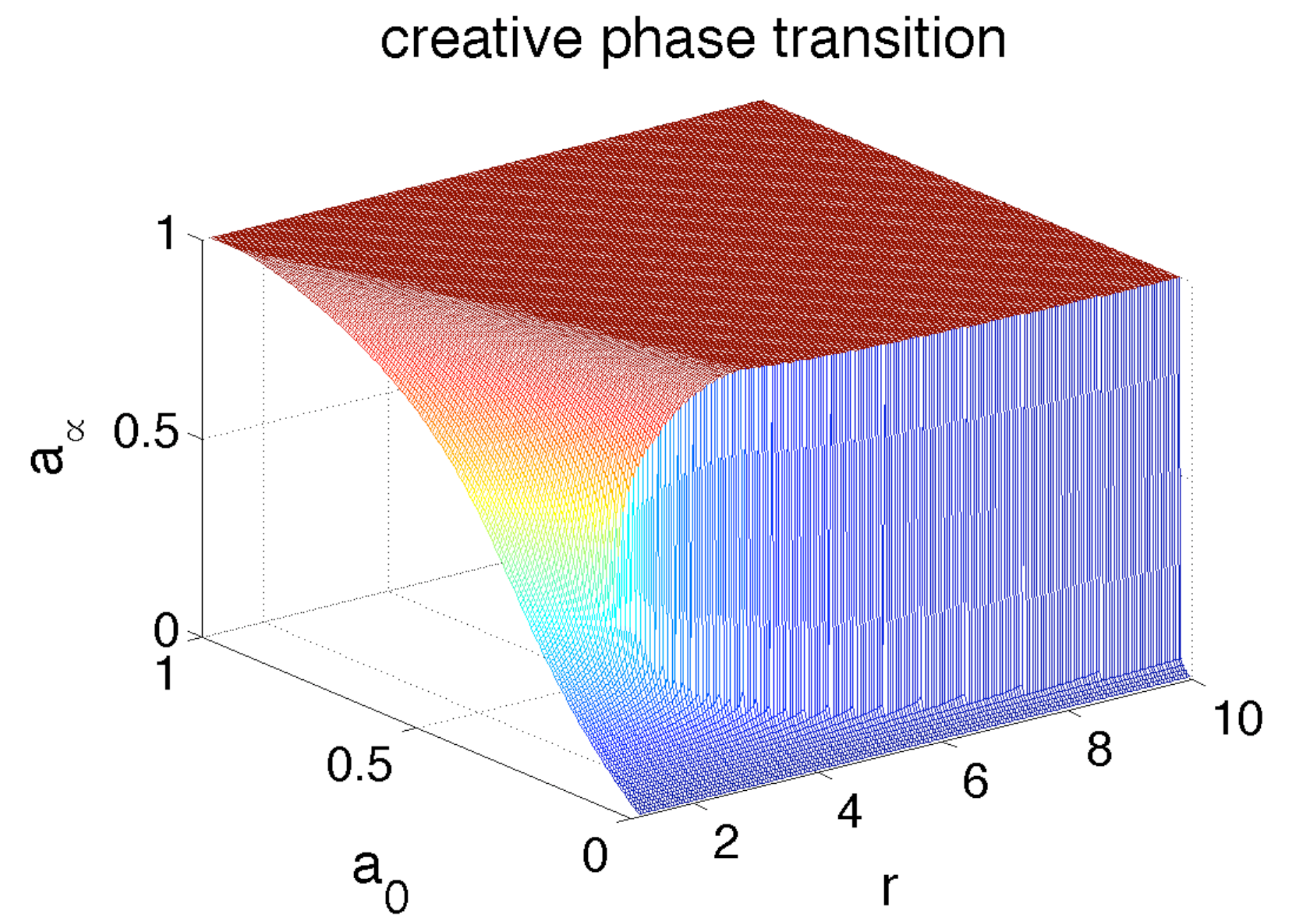}
\end{center}
\caption{Phase diagram of the creative dynamics over the $r$-$a_0$ space.}
\label{fig1}      
\end{figure}

If the number of non-zero elements in $x(t)$ is denoted by $a_t$, it was shown in  \cite{hanel05}
that the non-linear, second order recurrence equations associated with  Eq. (\ref{model}) are given by
\begin{equation}
 a_{t+1}=a_{t}+\Delta a_{t} \quad , \quad
 \Delta a_{t+1}=\frac{r}{d}\left(1-\frac{a_{t+1}}{d}\right)
 \left(a_{t+1}^{2}-a_{t}^{2}\right)   ,
\label{update}
\end{equation} 
with the initial conditions $a_{0}$ being the initial number of present elements 
and $a_{-1}\equiv0$, by convention. The question is to find the final diversity 
of the system, $a_{\infty}$. 
These equations are exactly solvable in the long-time limit. For this end 
define,  $c_{t}\equiv \Delta a_{t+1}/ \Delta a_{t}$, and look at the asymptotic behavior,  
$c \equiv \lim_{t\to\infty}c_t$.
From Eq. (\ref{update})  we get 
\begin{equation}
 c=2r\left(1-\frac{a_{\infty}}{d}\right)\frac{a_{\infty}}{d}  \quad. 
 \label{ansatz_a}
\end{equation} 
On the other hand we can estimate $a_{\infty}$ asymptotically by
\begin{equation}
 a_{\infty}= a_0 \sum_{t=0}^{\infty}c^t = \frac{a_0}{1-c} \quad .
\label{ansatz_b}
\end{equation} 
Introducing  Eq. (\ref{ansatz_a}) into Eq. (\ref{ansatz_b})
one gets a third order equation, whose solutions are the 
solution to the problem. Most remarkably these solutions 
are mathematically identical to the description of real gases, i.e. 
Van der Waals gases. As real gases our system shows a phase 
transition phenomenon. The corresponding phase diagram,  
as a function of the model parameter $r$ and the initial 
condition $a_0$ is shown in Fig. \ref{fig1}.

One can make the relation to the  Van der Waals gas more explicit by 
defining, $V \equiv a_{\infty}/d$ and $\tau  \equiv a_0/d$. 
Using this in Eqs. 
(\ref{ansatz_a}) and (\ref{ansatz_b}) 
gives 
$V-\tau=2r\left(1-V\right)V^2$.  
Renaming  variables
\begin{equation}
P\equiv \frac{1}{V}+\frac{1}{2rV^3} \qquad\mbox{and}\qquad T \equiv \frac{\tau}{2rV^3}
\label{variable_change}
\end{equation}   
leads to the famous equation, 
\begin{equation}
\left(P-\frac{1}{V^2}\right)V=T\quad, 
\label{vdwaals}
\end{equation}   
which is exactly a Van der Waals gas of point-particles with constant (negative) internal pressure.
The meaning of 'pressure' and 'temperature' in our context is intuitively clear.

\section{The destructive phase transition}

In the dynamics studied so far diversity can only increase due to the positivity of the  
elements in $\alpha$. It is important to note that in this setting 
the phase transition can not be crossed in the backward direction. 
This is because of two reasons. 
First, the system {\em forgets} its initial condition $a_0$ once it
has reached the (almost) fully populated state. This means that after everything has been 
produced one can not lower the initial set size any more. In terms of  the Van der Waals gas equation
analogy we can {\em not} lower the 'temperature' and we can {\em not} cross the phase 
transition in the backward direction.
Second, if $r$ is a homogeneous characteristic of the system then it is also impossible to manipulate 
the 'pressure' of the system and we remain in the fully populated phase for ever.

The natural question thus arises what happens to the dynamics if one 
randomly kills a fraction of elements in the fully (or almost fully) populated phase. 
In the case that an element $k$ gets produced by a single pair  $(i,j)$ and one of these 
-- either $i$ or $j$ --  
gets killed, $k$ can not be produced any longer. We call the random removal of $i$ 
a {\em primary defect}, the result -- here the stop of production of $k$ -- 
is a {\em secondary defect},  denoted by $SD$. 
The question is whether there exist critical conditions of $r$ and a  primary defect density $\delta_0$, 
such that cascading defects will occur. 

As before we approach this question iteratively, by asking how many secondary defects will be 
caused by an initial set of $D_0$ randomly removed elements in the  fully populated phase. 
We define the primary defect density $\delta_0\equiv D_0/d$.
The possibility for a secondary defect happening to element $k$ requires that {\rm all}  
productive pairs, which can produce $k$, have to be destroyed, i.e. 
at least one element of the productive pair has to be eliminated 
\footnote{On average  there are $r$ production pairs for $k$.}.
This requires some 'book-keeping' of the number of elements that partially 
have lost some of their productive pairs due to defects.
We introduce a book-keeping set $G_n$ of sequences $g_{nl}$, 
$G_n=\{g_{n0}, g_{n1}, g_{n2}, g_{n3}, \cdots\}$, where $d\,g_{nl}$ denotes the number of elements 
that have lost $l$ ways to be produced (i.e. productive pairs), given that initially 
$n$ elements have been eliminated. 

To be entirely clear, let us introduce the first defect. 
This defect will on average affect $2r$ productive pairs in the system, i.e.,  
there will be $2r$ elements that loose one way of being produced 
\footnote{Why? Since there are $d\,r$ productive  pairs there are $2\,d\,r$ indices referring to an 
element involved in denoting the pairs. Consequently there are $2\,r$ indices on average per element.}. 
We naturally assume $d\gg d\delta_0\gg r > 1$, 
and disregard the vanishingly small probability that one element looses two or 
more of its productive pairs by one primary defect. 

Before the first defect we have $G_0=\{1, 0, 0,\cdots\}$, meaning that there are $d$ entities that have lost none
of their producing pairs. The first defect will decrease this number $d\to d-2r$, 
i.e.,   we get $2r$ elements that have lost one of their producing pairs. 
Consequently we find $G_1=\{1-p, p, 0,0,\cdots\}$, where $p$ is defined as $p\equiv2r/d$.
Now, defecting the second element will affect another $2r$ elements through their 
producing pairs. This time  we affect an element that has lost none of its producing pairs 
with probability $1-p$,
and with probability $p$ we affect an element that already has lost one of its producing pairs. 
Iterating this idea of subsequent defects leads to the recurrence relations 
\begin{equation}
 g_{n+1\,0}=g_{n\,0}\,(1-p)\qquad  {\rm and } \qquad
 g_{n+1\,k}=g_{n\,k}+\left(g_{n\,k-1}-g_{n\,k}\right)\,p\quad .
 \label{recrel_g}
\end{equation}   
It is easy to show that  $g_{n\,k}$ follows a binomial law, 
$ g_{n\,k}=\binomial{n}{k}p^k(1-p)^{n-k} $. 
The number of secondary defects after $n$ introduced defects, denoted by $SD_n$, 
is just the number of all entities that 
have lost {\em all} of their (on average) $r$ producing pairs and can be estimated by $d\,\sum_{k\geq r} g_{n\,k}$.
Defining 
\begin{equation}
 SD_n=\sum_{k\geq r} g_{n\,k}\quad,
 \label{sec_definition}
\end{equation}   
one finds the update equation for $SD_n$ by inserting (\ref{recrel_g}) into (\ref{sec_definition}), 
\begin{equation}
 SD_{n+1}=SD_n+p g_{n\,r-1}\quad.
 \label{sec_recrel}
\end{equation}   
Now, if $d\,\delta_0$ and  $d\,\delta_1$ are the numbers of primary and secondary defects respectively, 
one has to identify
\begin{equation}
\delta_1=SD_{d \delta_0}\quad.
\label{sec_defects_a}
\end{equation}   
This is nothing but
\begin{equation}
\begin{array}{lll}
\delta_1&=&p\sum_{n\geq r}^{d\,\delta_0}g_{n\,r-1}
=p\sum_{n\geq r}^{d\,\delta_0} \binomial{n}{r-1}p^{r-1}(1-p)^{n-r+1}\quad.
\end{array}
\label{sec_defects_b}
\end{equation}   
Since we assume $d\gg d\delta_0\gg r > 1$,  Stirling's approximation is reasonable, 
$\ln(n!)\sim n\ln(n)-n +\frac{1}{2}\ln(2\pi n)$, 
so that the binomial coefficient is approximated by, 
$\binomial{n}{m}\sim\left(\frac{n}{m}\right)^m e^{m}(2\pi m)^{-1/2}$,
where $(1-m/n)^{n-m}\sim\exp(-m)$,  for $n\gg m$. Further one can approximate
$(1-p)^{n-r+1}\sim\exp(-np)$. Inserting these approximations into Eq. (\ref{sec_defects_b}), and replacing 
the sum by an integral one gets  
\begin{equation}
 \delta_1=\frac{p^r}{\sqrt{2\pi}}(r-1)^{\frac{1}{2}-r}e^{r-1}
 \int_{r}^{d\,\delta_0}\,dx\,x^{r-1}e^{-xp}\quad.
 \label{sec_defects_c}
\end{equation}   
Since $p\,d\delta_0=2\,r\,\delta_0$, and by approximating $p\,r\sim 0$ (for the lower limit) we rewrite the integral 
\begin{equation}
 \int_{r}^{d\,\delta_0}\,dx\,x^{r-1}e^{-xp}\sim p^{-r}\int_{0}^{2r\delta_0}\,dy\,y^{r-1}e^{-y}\quad,
 \label{rewrite_integral}
\end{equation}   
and we can finally compute
\begin{equation}
 \delta_1=\gamma(r)f(r,\delta_0)\delta_0^{r}\quad,
 \label{sec_defects_d}
\end{equation}   
with 
\begin{equation}
 \begin{array}{lll}
     \gamma(r)&=&\frac{1}{r}\frac{(2r)^r}{\sqrt{2\pi}}(r-1)^{\frac{1}{2}-r}e^{r-1}\\&&\\
  f (r,\delta_0)&=&\sum_{n=0}^{\infty}\frac{1}{n!}\frac{r}{r+n}(-2r\delta_0)^n\quad. 
 \end{array}
 \label{sec_defects_ad_d}
\end{equation}   
Here $f$ is obtained by expanding the exponential in the integral of Eq. (\ref{rewrite_integral}) into a Taylor series.

What remains to be done is to iterate Eq. (\ref{sec_defects_d}). 
There are two possible ways of doing so.
In the first iteration scheme we think of collecting the primary and secondary defects together and
assume that we would start with a new primary defect set of size $\delta_0'=\delta_0+\delta_1$. 
The tertiary defects therefore would be estimated by 
$\delta_2=\delta_1'-\delta_1$, 
where $\delta_1'$ are the secondary defects associated with $\delta_0'$. This leads to the recursive scheme (A), 
\begin{equation}
\Delta_n\equiv\sum_{k=1}^n\delta_k \quad,\qquad 
\delta_{n+1} = \gamma(r)f(r,\Delta_n)\Delta_n^r -\Delta_n + \delta_1\quad ({\rm A } ).
\label{defect_recrel_A}
\end{equation}

The second way to iterate Eq. (\ref{sec_defects_d}) is to assume that we use the $d\delta_1$ secondary defects
as primary defects on the smaller (rescaled) system $d(1-\delta_0)$ so that we look at a new primary defect-ratio
$\delta_0'=\delta_1/(1-\delta_0)$. The result $\delta_1'$ then has to be rescaled inversely to give
the tertiary defects in the original scale, i.e. $\delta_2=(1-\delta_0)\delta_1'$.
Iterating this idea leads to the recurrence relation (B),
\begin{equation}
 \Delta_n\equiv\sum_{k=1}^n\delta_k \quad,\qquad 
 \delta_{n+1} = \gamma(r)\left(1-\Delta_{n-1}\right)^{1-r}f\left(r, \frac{\delta_n}{1-\Delta_{n-1}}\right)\delta_n^r \quad ({\rm B } ),
 \label{defect_recrel_B}
\end{equation}
with $\Delta_0\equiv 0$. 

\begin{figure}
\begin{tabular}{cc}
\includegraphics[width=7cm]{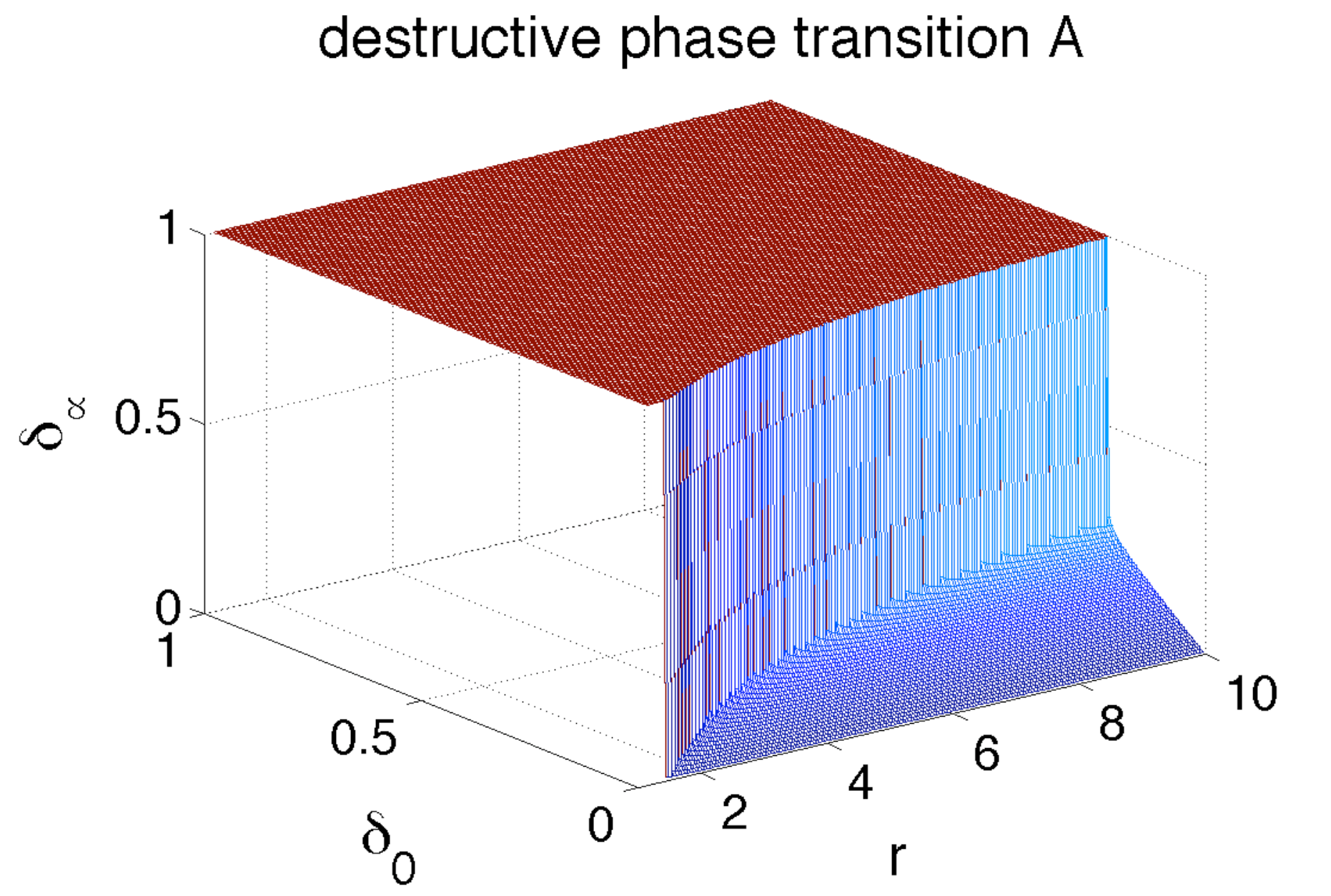} &
\includegraphics[width=7cm]{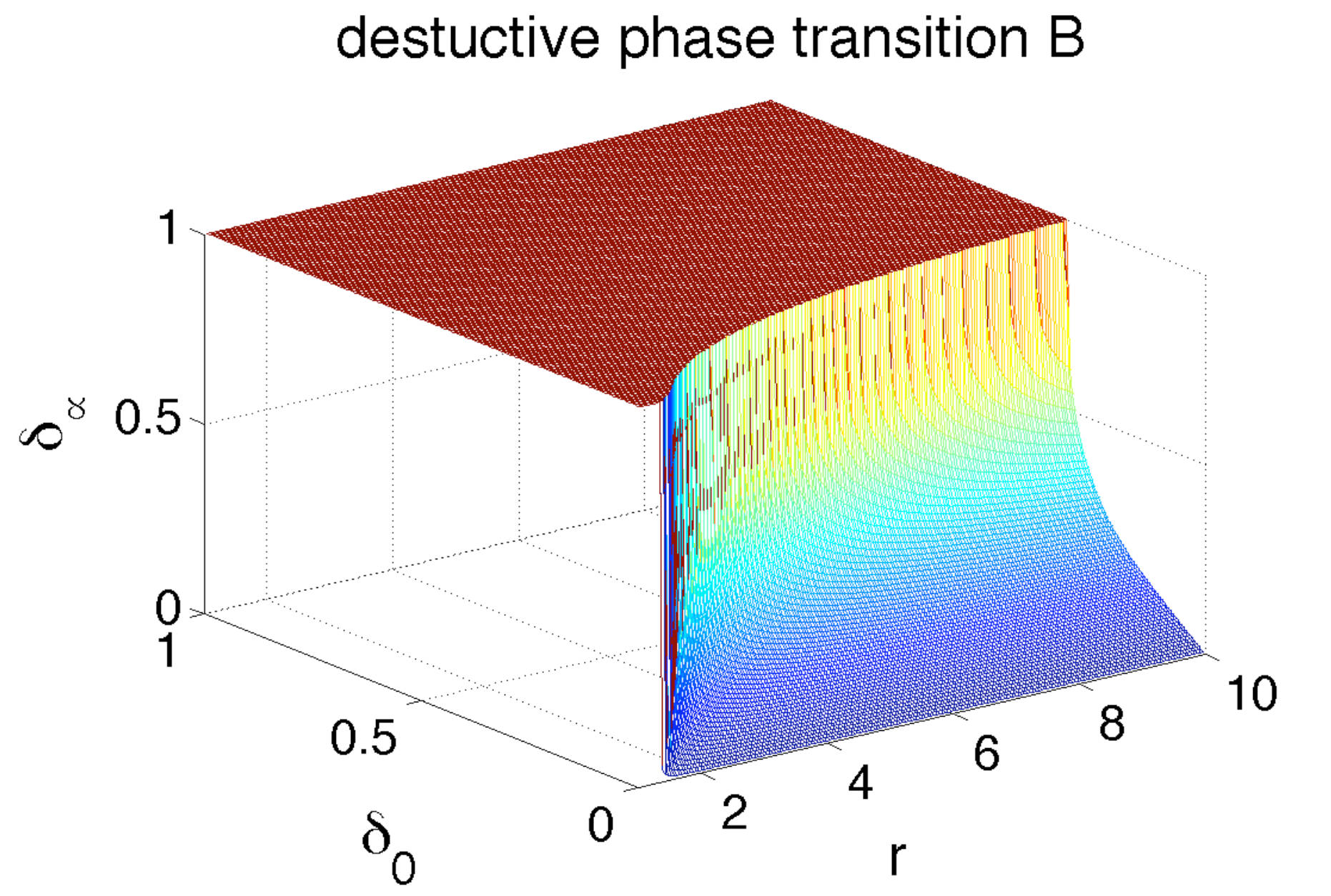} 
\end{tabular}
\caption{Phase diagram for the defect dynamics for two ways of iterating  Eq. (\ref{sec_defects_d})
described in the text. 
}
\label{fig2}      
\end{figure}

The result in terms of a phase diagram of the two possible iteration schemes (A) and (B) is 
given in Fig. \ref{fig2} (a) and (b), 
respectively. The asymptotic defect size $\delta_{\infty }$ (for $t\to \infty$) is shown as a function of the 
parameters $r$ and the initial defect density $\delta_0$. As before 
a clear phase transition is visible, meaning that at a fixed value of $r$ there exists a critical 
number of initial defects at which the system will experience a catastrophic decline of diversity. 
Unfortunately,  an analytical solution for the asymptotic iterations of Eq. (\ref{sec_defects_d})
seems to be beyond the capabilities of the authors. 
It is interesting that complete destruction of diversity (plateau in Fig. \ref{fig2}) not very large values 
of $\delta_0$ are necessary.

\begin{figure}
\begin{tabular}{cc}
\includegraphics[width=7cm]{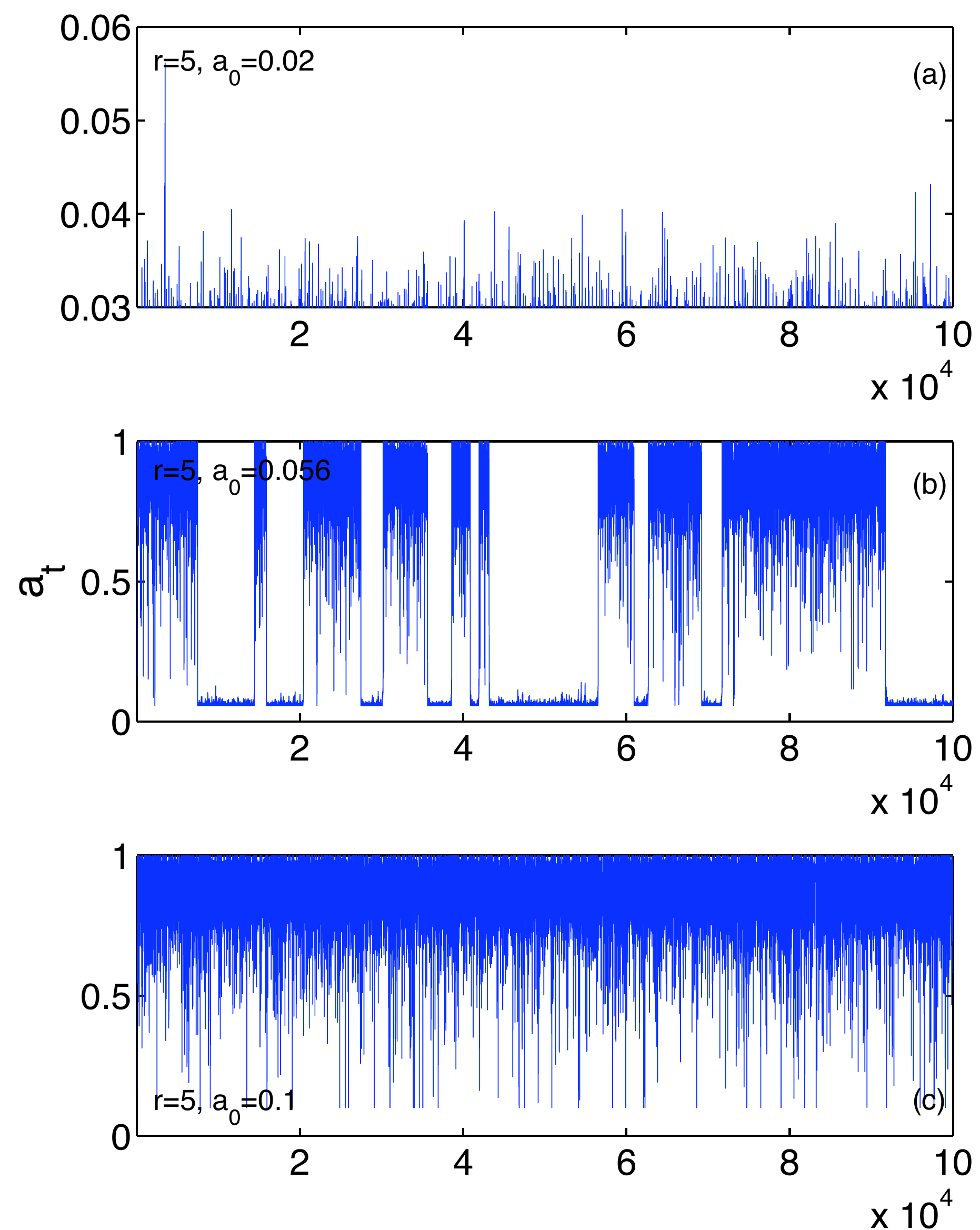} 
\includegraphics[width=7cm]{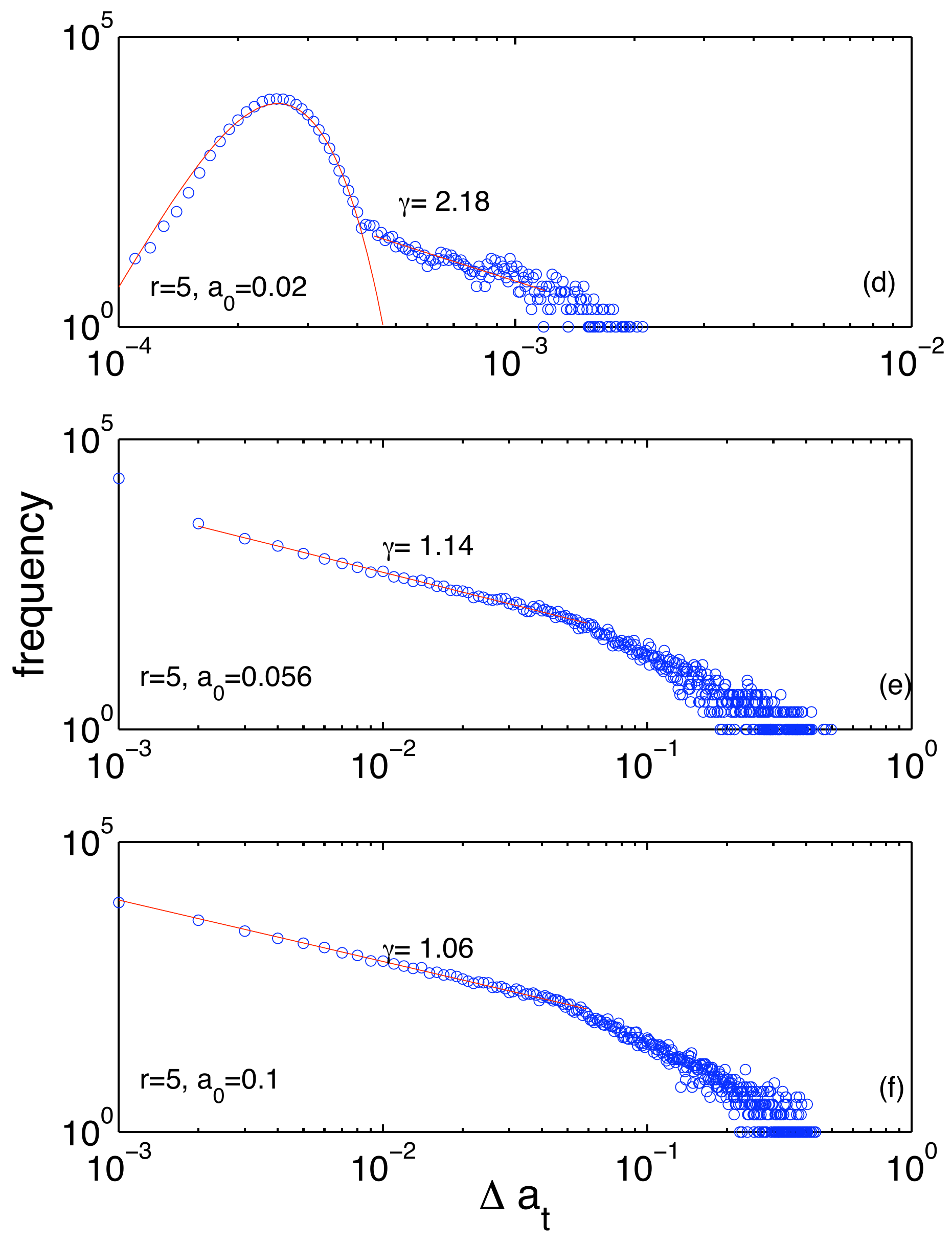} 
\end{tabular}
\caption{Time series and  time increment distribution of the coupled  dynamics for a 
fixed $r=5$ and fixed Poissonian rates  $\eta$ and  $\chi^{\pm }$. The variable varied is 
 $a_0=0.02,0.056$, and $0.1$. Red lines are fits to power laws with slopes $\gamma$, 
and a Poissonian resembling the creative driving noise in   (d). Note the change of scale 
here.}
\label{fig3}      
\end{figure}

\section{Combined dynamics: Creative gales of deconstruction}

To become more realistic, 
since we have now established the existence of phase transitions in both the creative and 
destructive regimes, and are equipped with the update equations for the respective 
cases Eqs.  (\ref{update}) and (\ref{sec_defects_d}), it is natural to couple these update equations 
and to study the combined dynamics.
The relevant variable now becomes the diversity in the system as a function of time, $a_t$.  
However, the question how this should be done is neither trivial nor uniquely determined. 

One realistic scenario might be that  at any point in time some goods/species/elements may 
come into being 
spontaneously and others go extinct at certain rates. First, for the introduction of new elements 
we introduce a stochastic rate,  
$\chi^+>0$ of a Poisson process,  so that  $(d-a_t)\chi^+_t$ new species may 
be expected in one time unit. Note, that there are $d-a_t$ 'un-populated' elements in the system. 
These randomly created elements are elements that did not get produced through 
(re)combination or substitution of existing ones, but are 'out of the blue' inventions. 
The natural time unit we are supplied with is one creative generation 
$a_t\to  a_{t+1}$. The spontaneous creation may eventually increase the critical threshold 
and the system may transit into the highly diverse phase (think of this process to 
randomly alter $a_0$ in the creative update dynamics).

On the other hand there are  spontaneous processes that destroy or 
remove species  at a stochastic rate, $\chi^->0$ (Poisson process), such that about $a_t\chi^-_t$ 
new defects may be expected per time unit. 
It can not be assumed a priori  that the iterative accumulation of secondary defects in the system, 
as described above, operates at the same time scale as the spontaneous or the deterministic creative processes.

For making an explicit choice we may assume that during one time unit there  happen 
$\eta_t$ generations of secondary defects, taking into account 
the relative ratio of innovative and secondary defect generations processed
per time unit. We assume that $\eta_t$ can be modeled by a Poisson process whose rate, $\left<\eta_t\right>=\eta$
becomes  a parameter of the model. 
For the computations below we have chosen $\eta=0.1$. 

When we look at the way secondary defects evolve in generations we are left with a 
culminated number of secondary defects $\Delta_{\eta_t\,t}$ after $\eta_t$ generations and a remainder 
$\delta_{\eta_t\,t}$, which would have to be added to $\Delta_{\eta_t\,t}$ in the next defect-generation,  
$\eta_t+1$ 
but which -- by assumption -- is falling into the 
book-keeping of the next creative-generation time step $t+1$. 
What we say is that 
during time step $t\to t+1$, there are $\Delta^- a_{t}=d \Delta_{\eta_t,t}$ 
species removed from the system, where $\Delta_{m\,t}=\sum_{k=0}^{m-1}\delta_{k\,t}$ is the 
cumulated ratio of secondary defect ratios $\delta_{k\,t}$ of defect-generation $k$ at time step $t$. 
The remaining defects of generation $\eta_t$ have to be accounted for in the next time 
step together with the newly introduced spontaneous defects, so that 
$\delta_{0\,t+1} = \frac{a_{t}}{d}\chi^-_t + \delta_{\eta_t,t}$. The update of defect generations 
now can be performed  $\eta_t$ times according to 
\begin{equation}
\delta_{m+1\, t}=\left(1-\Delta_{m\, t}\right)\gamma_r f(r, \tilde\delta_{m\, t})\tilde\delta_{m\, t}^r\quad,
\end{equation}
where we have considered the rescaling approach (B) to secondary defect generations. 
A similar equation can be derived for scheme (A).
For convenience of notation we write for the rescaled defect ratios, 
$\tilde\delta_{m\,t} \equiv \frac{\delta_{m\,t}}{1-\Delta_{m\,t}}$. If 
now, by coincidence, the remaining defects from the last time step and the spontaneously 
introduced defects are sufficiently many and there are enough defect-generations $\eta_t$ 
processed in that time step, the culminating secondary defects may lead to a break down 
of the system from the high  to the low  diversity regime.

All that is left is to insert this dynamics into the creative update equation.
To do so we first note that without defects, $\Delta a_t$ depends on both $a_t$ and $a_{t-1}$.
However, due to the occurring defects $a_{t-1}$ will not remain what it was when $t$ becomes updated
to $t+1$, but will be decreased by the occurring defects in this time span. 
For this reason it is convenient to introduce a new variable $b_t$ which  takes the place of
$a_{t-1}$ in the coupled update process. More precisely,
$b_{t+1} \equiv a_{t}- \Delta^- a_{t}$. 
For the growth condition to be well defined we require $a_t>b_t$, which is guaranteed by
$a_{t+1} = b_{t+1}+ \Delta^+ a_{t}$ where
\begin{equation}
ÊÊÊÊÊÊÊ \Delta^+ a_{t} \equiv \frac{r}{d}\left(1-\frac{a_{t}}{d}\right)\left(a_{t}^{2} - b_{t}^{2}\right)+(d-a_{t})\chi^+_t \quad ,
\end{equation}
is the number of deterministically (by the creative update law) and spontaneously 
introduced species in the creative-generation $t$.
This sort of coupling allows to take a look at how diversity of systems may evolve over time, 
driven by the spontaneous creation and destruction processes $\chi^\pm$, which may reflect 
exogenous influences,
 while on the other hand the average number of defect-generations $\eta$ per creative generation 
$t$, and the average number of productive pairs per species $r$  
express endogenous properties of the system, i.e. whether the defects process 
slow or fast ($\eta$),  and the average dependency ($r$) of the catalytic network.

We study the resulting timeseries for this dynamics for several values of $r$, $a_0$, $\eta$, and  $\chi^{\pm }$.
In Fig. \ref{fig3}, by  fixing $r=5$ and the Poisson rates $\eta$, and  $\chi^{\pm }$ and by varying 
$a_0$ from 0.01 to 0.1, we cross the creative phase transition line from the sub-critical to the fully populated phase. 
At $a_0=0.056$ we observe a flip-flop transition between the two phases.   
The flip-flop transitions happen over very short time intervals. 
In Fig. \ref{fig3} (b) the increment distribution of $\Delta a_t\equiv a_t-a_{t-1}$ is shown. It is 
clearly seen that the distribution is power law whenever the super-critical phase is sufficiently populated. 
The Poissonian driving in the creative dynamics in the sub-critical 
region is clearly seen for $a_0=0.002$ in Fig. \ref{fig3} (d).  
By power-law fits to the exponents in the deconstructive regime, we observe a sign for an existence of 
a Zipf law, i.e. $\gamma\sim 1$.

\section{Conclusion}

We have shown the existence of a new phase transition in systems capable of evolutionary dynamics.
Given that the system is in its highly diverse state, the introduction of relatively little primary removal 
of elements can cause drastic declines in diversity. We have further proposed a dynamical model to 
study timeseries of diversity in systems governed by the evolution equation (\ref{model}) under the influence of 
external spontaneous creation and destruction processes. 
We emphasize that we strictly stick to the structure of Eq. (\ref{model}) and do not discuss variants, such as the 
beautiful work of \cite{jain}. In contrast they have studied a linear version (resembling catalytic equations), 
however with an explicit 'selection' mechanism incorporated in a dynamic rule table. 

As the main result of this present work we re-discover  what J.A. Schumpeter has heuristically and 
qualitatively described as creative gales of deconstruction. 
More importantly we are able to quantify the dynamics of such systems. As an example 
destructive processes can be quantified in real world situations by bankruptcies of firms. 
In this context  the existence of a power law and in particular empirical evidence for a Zipf law 
-- similar to the one resulting from our model -- has been found in \cite{zipf}. 
As in the work of \cite{jain} we observe  the importance of different time scales in the coupled dynamics. 
In our approach we have incorporated this aspect by noting that creation and destruction 
do not work necessarily on the same time scales. 
Let us mention as a final comment that the results do of course not only apply to technological evolution but 
to any biological, social, or physical system governed by the evolution equation, Eq.  (\ref{model}). 

\bibliographystyle{unsrt}

\end{document}